\documentclass[a4paper,english,aps, prb, longbibliography, unsortedaddress]{revtex4-1}
\usepackage[T1]{fontenc}
\usepackage[latin9]{inputenc}
\usepackage{babel}
\usepackage{amsmath}
\usepackage{amssymb}
\usepackage{graphicx}
\usepackage[unicode=true,pdfusetitle,
 bookmarks=true,bookmarksnumbered=false,bookmarksopen=false,
 breaklinks=false,pdfborder={0 0 1},backref=false,colorlinks=false]
 {hyperref}

\makeatletter

\providecommand{\tabularnewline}{\\}

\makeatother

\begin{document}
\title{Theoretical analysis of thermal boundary conductance of MoS$_{2}$-SiO$_{2}$
and WS$_{2}$-SiO$_{2}$ interface}
\author{Zhun-Yong Ong }
\email{ongzy@ihpc.a-star.edu.sg}

\affiliation{Institute of High Performance Computing, A{*}STAR, Singapore 138632,
Singapore}
\author{Yongqing Cai }
\affiliation{Joint Key Laboratory of the Ministry of Education, Institute of Applied
Physics and Materials Engineering, University of Macau, Taipa, Macau,
China}
\author{Gang Zhang }
\affiliation{Institute of High Performance Computing, A{*}STAR, Singapore 138632,
Singapore}
\author{Yong-Wei Zhang}
\affiliation{Institute of High Performance Computing, A{*}STAR, Singapore 138632,
Singapore}
\date{\today}
\begin{abstract}
Understanding the physical processes involved in interfacial heat
transfer is critical for the interpretation of thermometric measurements
and the optimization of heat dissipation in nanoelectronic devices
that are based on transition metal dichalcogenide (TMD) semiconductors.
We model the phononic and electronic contributions to the thermal
boundary conductance (TBC) variability for the MoS$_{2}$-SiO$_{2}$
and WS$_{2}$-SiO$_{2}$ interface. A phenomenological theory to model
diffuse phonon transport at disordered interfaces is introduced and
yields $G=13.5$ and $12.4$ MW/K/m$^{2}$ at 300 K for the MoS$_{2}$-SiO$_{2}$
and WS$_{2}$-SiO$_{2}$ interface, respectively. We compare its predictions
to those of the coherent phonon model and find that the former fits
the MoS$_{2}$-SiO$_{2}$ data from experiments and simulations significantly
better. Our analysis suggests that heat dissipation at the TMD-SiO$_{2}$
interface is dominated by phonons scattered diffusely by the rough
interface although the electronic TBC contribution can be significant
even at low electron densities ($n\leq10^{12}$ cm$^{-2}$) and may
explain some of the variation in the experimental TBC data from the
literature. The physical insights from our study can be useful for
the development of thermally aware designs in TMD-based nanoelectronics.
\end{abstract}
\maketitle

\section{Introduction}

Atomically thin two-dimensional (2D) transition metal dichalcogenide
(TMD) semiconductors such as MoS$_{2}$ and WS$_{2}$ hold great potential
for the development of next-generation electronic devices~\citep{XLi:AdvMater15_Performance,DLembke:ACR15_Single}.
At the nanoscale, high power densities in these devices require efficient
thermal management crucial for optimal device performance, with the
thermal boundary conductance (TBC) of the 2D crystal-substrate interface
playing a key role in the dissipation of excess Joule heat~\citep{EPop:NR10_Energy,ZYOng:2DM19_Energy}.
Therefore, clearer insights into the role of the different physical
mechanisms underlying the TBC of the TMD-substrate interface may lead
to superior thermally aware TMD-based nanoelectronic device designs. 

One widely studied mechanism is the van der Waals coupling between
the phonons of the 2D crystal and its substrate which is believed
to be the dominant component in the overall TBC~\citep{ZYOng:2DM19_Energy}.
The phononic TBC ($G_{\text{ph}}$) has been estimated using molecular
dynamics (MD) simulations~\citep{ZYOng:JAP18_Flexural,SVSuryavanshi:JAP19_Thermal},
elasticity theory~\citep{BNJPersson:JPCM11_Phononic,ZYOng:PRB16_Theory,ZYOng:PRB17_Thickness}
and density functional theory-based models~\citep{GCCorrea:Nanotech17_Interface}.
Another mechanism of heat dissipation is through the inelastic scattering
of electrons in the 2D crystal by dipoles in the dielectric substrate,
a phenomenon known widely as ``remote phonon (RP) scattering'' or
``surface optical phonon scattering''~\citep{KHess:SSC79_Remote,MVFischetti:JAP01_Effective,AKonar:PRB10_Effect,KZou:PRL10_Deposition}.
This inelastic scattering mechanism, which plays an important role
in limiting the electron mobility in TMDs~\citep{ZYu:NatCommun14_Towards,ZYu:AdvMater16_Realization},
also underlies the electronic TBC ($G_{\text{el}}$) and depends strongly
on the electron or hole density. Although it is predicted to be insubstantial
for heat dissipation in graphene~\citep{ZYOng:PRB13_Signatures,YKoh:NL16_Role},
the contribution from RP scattering may be significant and comparable
to the phononic TBC for MoS$_{2}$ and WS$_{2}$~\citep{ZYOng:Manuscript20_Remote}. 

Nonetheless, even though the phononic TBC has been studied using various
theoretical methods, it is useful to have a model that relates $G_{\text{ph}}$
to the fundamental elastic properties of the 2D crystal and its substrate
while accounting for quantum statistics at low temperatures. One such
theory that is based on coherent flexural phonons~\citep{BNJPersson:JPCM11_Phononic}
and developed in Ref.~\citep{ZYOng:PRB16_Theory} yields relatively
accurate values for the graphene-SiO$_{2}$ interface~\citep{ZYOng:PRB16_Theory,ZYOng:PRB17_Thickness}
and also predicts a modest room-temperature phononic TBC of $\sim3.1$
MW/K/m$^{2}$ for the MoS$_{2}$-SiO$_{2}$ interface, in good agreement
with earlier published data~\citep{ATaube:ACSApplMaterInterf15_Temperature}
but considerably smaller than later more accurately measured TBC values~\citep{EYalon:ACSApplMaterInterf17_Temperature,EYalon:NL17_Energy}
in the range of $10$ to $15$ MW/K/m$^{2}$. This substantial discrepancy
necessitates a relook of the heat dissipation mechanisms at the TMD-substrate
substrate.

In our paper, we model the relative contribution of the different
physical mechanisms to the overall TBC of the MoS$_{2}$-SiO$_{2}$
and WS$_{2}$-SiO$_{2}$ interface. Our objective is to understand
how these mechanisms (e.g. coherent vs. diffuse phonon transport and
the electron density dependence of $G_{\text{el}}$) lead to the variability
in their TBC. We revisit the phononic TBC problem and introduce a
phenomenological \emph{diffuse} phonon theory that is more suitable
for atomically disordered interfaces and allows us to model the effect
of disorder on the TBC. To understand its difference to the coherent
theory, we compare the TBC predictions from the coherent and diffuse
phonon models with data from experiments and MD simulations. We also
compare the phononic contribution (coherent or diffuse) to the electronic
contribution in the TBC. Finally, we use the combined electronic and
phononic BC results ($G=G_{\text{el}}+G_{\text{ph}}$) to analyze
reported experimental TBC data~\citep{ATaube:ACSApplMaterInterf15_Temperature,EYalon:NL17_Energy,EYalon:ACSApplMaterInterf17_Temperature,PYasaei:AdvMaterInterf17_Interfacial}
for the MoS$_{2}$-SiO$_{2}$ interface and to discuss the possible
physics underlying the variability of the TBC data. We also use the
combined electronic and phononic BC results to predict the variability
of the TBC for the WS$_{2}$-SiO$_{2}$ interface. Although the theoretical
methods in this paper are used specifically to analyze the TBC variability
of the single-layer MoS$_{2}$-SiO$_{2}$ and WS$_{2}$-SiO$_{2}$
interface, they can also be extended to other 2D crystals (e.g. graphene
and other TMDs) and substrates (e.g. Al$_{2}$O$_{3}$). 

\section{Theoretical models}

\begin{figure}
\begin{centering}
\includegraphics[width=8cm]{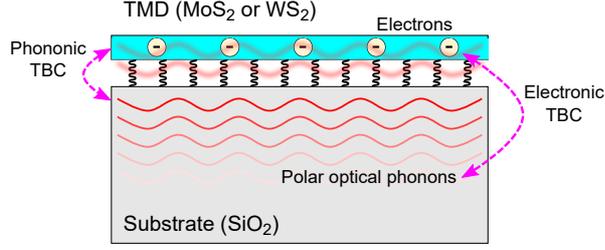}
\par\end{centering}
\caption{Schematic of the TMD-substrate interface and the phononic and electronic
heat dissipation processes (represented by dashed lines). The TMD
sheet is mechanically coupled to the SiO$_{2}$ substrate through
the effective spring forces (represented by vertical wavy lines) at
the interface while the TMD electrons (represented by circles) are
coupled to the polar optical phonons (represented by horizontal wavy
lines) in the substrate.}
\label{fig:InterfaceStructureSchematic}
\end{figure}

We discuss our theoretical models of the phononic and electronic TBC
processes, as depicted in the schematic of the TMD-substrate interface
in Fig.~\ref{fig:InterfaceStructureSchematic}, in the following
subsections. Roughly speaking, we attribute the phononic TBC ($G_{\text{ph}}$)
to the linear mechanical coupling between the flexural phonons of
the TMD sheet and the bulk elastic waves (i.e., acoustic phonons)
of the substrate, and the electronic TBC ($G_{\text{el}}$) to the
remote scattering of the TMD electrons by the polar optical phonons
of the dielectric substrate, also otherwise known as ``remote phonon
scattering''~\citep{KHess:SSC79_Remote,ZYOng:Manuscript20_Remote}.
We assume that the two TBC components are independent (i.e., the electron-phonon
interaction has no effect on the mechanical coupling between the TMD
and the substrate and vice versa) and can be added in parallel. 

\subsection{Heat dissipation from coherent and diffuse flexural phonon scattering}

We adopt a linear elasticity theory-based approach to model the phononic
TBC from Ref.~\citep{ZYOng:PRB16_Theory}. Although the model is
applied to the TMD-SiO$_{2}$ interface here, it is sufficiently general
to be applied to any linear elastic isotropic substrate. The model,
which assumes heat dissipation by coherent 2D flexural phonons at
a \emph{perfectly smooth interface}~\citep{BNJPersson:JPCM11_Phononic},
is used to derive the following expression for the \emph{coherent}
phononic TBC 
\begin{equation}
G_{\text{ph}}^{\text{coh}}=\frac{1}{2\pi}\int_{0}^{\infty}d\omega\ \hbar\omega\frac{dN(\omega,T)}{dT}\Xi_{\text{coh}}(\omega)\ ,\label{eq:CoherentPhononTBC}
\end{equation}
where 
\begin{equation}
\Xi_{\text{coh}}(\omega)=\frac{1}{(2\pi)^{2}}\int_{q<q_{\text{max}}}d^{2}q\frac{4K^{2}\text{Im}D_{\text{sub}}(\boldsymbol{q},\omega)\text{Im}D_{\text{2D}}(\boldsymbol{q},\omega)}{|1-K[D_{\text{sub}}(\boldsymbol{q},\omega)+D_{\text{2D}}(\boldsymbol{q},\omega)]|^{2}}\label{eq:Transmission_coherent}
\end{equation}
is the \emph{coherent} areal transmission function, $q_{\text{max}}$
is the maximum wave vector which we can set as $q_{\text{max}}=2\pi/\sqrt{A}\approx2\times10^{10}$
m$^{-1}$ ($A$ is the unit cell area of the TMD), and $N(\omega,T)=[\exp(\frac{\hbar\omega}{k_{B}T})-1]^{-1}$
is the Bose-Einstein occupation factor at frequency $\omega$ and
temperature $T$. The transmission function in Eq.~(\ref{eq:Transmission_coherent})
depends on the spring constant at the TMD-substrate interface $K$,
the retarded Green's function for the flexural motion of the TMD monolayer
$D_{\text{2D}}(\boldsymbol{q},\omega)$, and the retarded Green's
function for the free surface displacement of the isotropic solid
substrate $D_{\text{sub}}(\boldsymbol{q},\omega)$. The value of $K$
for the MoS$_{2}$-SiO$_{2}$ interface is taken from Ref.~\citep{ZYOng:PRB16_Theory}
while its value for the WS$_{2}$-SiO$_{2}$ interface is calculated
using density functional theory like in Ref.~\citep{ZYOng:PRB16_Theory}. 

The expressions for $D_{\text{2D}}(\boldsymbol{q},\omega)$ and $D_{\text{sub}}(\boldsymbol{q},\omega)$
are~\citep{ZYOng:PRB16_Theory}
\begin{equation}
D_{\text{2D}}(\boldsymbol{q},\omega)=\lim_{\eta\rightarrow0^{+}}[\rho\omega^{2}+i\rho\gamma(\omega)\omega-\kappa q^{4}+i\eta]^{-1}\label{Eq:2DFlexuralGreensFunction}
\end{equation}
where $\rho$ and $\kappa$ denote the areal mass density and the
bending stiffness of the uncoupled 2D crystal, respectively, and
\begin{equation}
D_{\text{\text{sub}}}(\boldsymbol{q},\omega)=\frac{i}{\rho_{\text{sub}}c_{T}^{2}}\frac{p_{L}(q,\omega)}{S(q,\omega)}\left(\frac{\omega}{c_{T}}\right)^{2}\Theta(\omega_{D}-\omega)\label{Eq:SubstrateGreensFunction}
\end{equation}
where \begin{subequations} 
\begin{align}
S(q,\omega) & =\left[\left(\frac{\omega}{c_{T}}\right){}^{2}-2q^{2}\right]{}^{2}+4q^{2}p_{T}p_{L}\ ,\\
p_{L}(q,\omega) & =\lim_{\eta\rightarrow0^{+}}\left[\left(\frac{\omega}{c_{L}}\right)^{2}-q^{2}+i\eta\right]^{1/2}\ ,\\
p_{T}(q,\omega) & =\lim_{\eta\rightarrow0^{+}}\left[\left(\frac{\omega}{c_{T}}\right)^{2}-q^{2}+i\eta\right]^{1/2}\ ,
\end{align}
\end{subequations}and $c_{L}$, $c_{T}$ and $\rho_{\text{sub}}$
are the longitudinal and transverse speed of sound, and the voluminal
mass density of the substrate, respectively. In Eq.~(\ref{Eq:SubstrateGreensFunction}),
$\Theta(\ldots)$ represents the Heaviside function, and $\omega_{D}=(6\pi^{2}N_{\text{sub}})^{1/3}c_{L}=9.39\times10^{13}$
rad/s (or $61.8$ meV) and $N_{\text{sub}}=6.62\times10^{28}$ m$^{-3}$
are respectively the longitudinal Debye frequency and number density~\citep{KGoodson:JHT94_Prediction}
of the substrate (amorphous SiO$_{2}$). As in Ref.~\citep{ZYOng:PRB16_Theory},
$\gamma(\omega)$ in Eq.~(\ref{Eq:2DFlexuralGreensFunction}) is
the frequency-dependent damping coefficient representing the \emph{intrinsic}
damping of the flexural motion, given as $\gamma(\omega)=\frac{\omega T}{\alpha T_{\text{RT}}}$
where $\alpha$ is the ratio of the phonon lifetime to its period
at room temperature and $T_{\text{RT}}$ is the room temperature ($300$
K). It has been shown~\citep{ZYOng:PRB16_Theory} that it is necessary
to include the intrinsic damping of the flexural motion in Eq.~(\ref{Eq:2DFlexuralGreensFunction})
to account for the interfacial heat dissipation correctly. 

Although the predictions of the coherent flexural phonon model are
relatively accurate for graphene~\citep{ZYOng:PRB16_Theory,ZYOng:PRB17_Thickness},
its predictions for the MoS$_{2}$-SiO$_{2}$ interface are substantially
smaller than experimental data~\citep{EYalon:ACSApplMaterInterf17_Temperature,EYalon:NL17_Energy},
highlighting a possible shortcoming of the model. The theory in Eq.~(\ref{eq:Transmission_coherent})
assumes a perfectly smooth interface~\citep{BNJPersson:JPCM11_Phononic}
which limits energy transfer to processes that conserve parallel momentum,
a restriction that may not apply to the more disordered TMD-substrate
interface which can have surface roughness, trapped charged impurities
and surface adsorbates~\citep{DRhodes:NatMater19_Disorder}. For
a highly disordered interface, this assumption can be relaxed and
we may assume that each incoming phonon is scattered by the disordered
interface without constraints to all available outgoing phonon modes
with \emph{equal probability} like in the diffuse mismatch model (DMM)~\citep{ETSwartz:RMP89_Thermal}.
Using this assumption, we apply the DMM approximation to modify Eq.~(\ref{eq:CoherentPhononTBC})
by writing the diffuse transmission function between the 2D crystal
and the substrate as~\citep{JZhang:JHT18_Optimizing}
\begin{equation}
\Xi_{\text{diff}}(\omega)=\frac{M_{\text{2D}}(\omega)M_{\text{sub}}(\omega)}{M_{\text{2D}}(\omega)+M_{\text{sub}}(\omega)}\ ,\label{eq:Transmission_DMM}
\end{equation}
where $M_{\text{2D}}(\omega)$ and $M_{\text{sub}}(\omega)$ are the
number of available transmitting modes per unit area at frequency
$\omega$ in the 2D crystal and substrate, respectively. We stress
that Eq.~(\ref{eq:Transmission_DMM}) describes a \emph{phenomenological}
model derived from purely kinetic considerations of detailed balance~\citep{ETSwartz:RMP89_Thermal,JZhang:JHT18_Optimizing}
and ignores the details of the interactions at the interface. The
corresponding TBC is
\begin{equation}
G_{\text{ph}}^{\text{diff}}=\frac{1}{2\pi}\int_{0}^{\infty}d\omega\ \hbar\omega\frac{dN(\omega,T)}{dT}\Xi_{\text{diff}}(\omega)\ .\label{eq:TBC_diffuse_limit}
\end{equation}

\begin{figure}
\begin{centering}
\includegraphics[width=11cm]{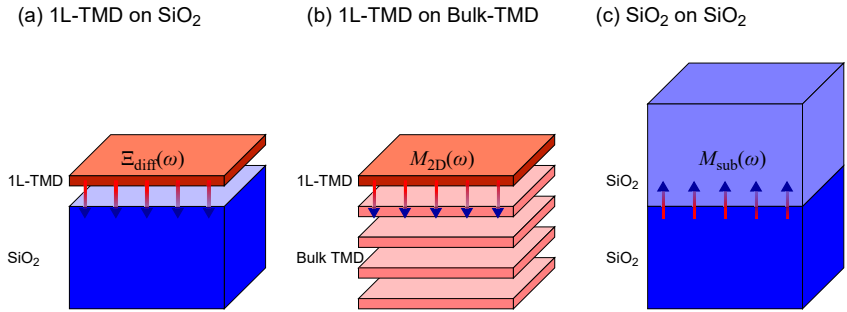}
\par\end{centering}
\caption{Schematic of the interfacial configurations used to determine (a)
the diffuse phonon transmission function $\Xi_{\text{diff}}(\omega)$
between a 1-layer TMD crystal (`1L-TMD') and a SiO$_{2}$ substrate
, (b) the number of modes $M_{\text{2D}}(\omega)$ between a 1L-TMD
and a bulk TMD substrate, and (c) the number of modes $M_{\text{sub}}(\omega)$
between two semi-infinite SiO$_{2}$ solids.}
\label{fig:DiffuseTransmissonSchematics}
\end{figure}

To determine $\Xi_{\text{diff}}(\omega)$ in Eq.~(\ref{eq:Transmission_DMM}),
we need to count the number of modes in the 2D crystal and the substrate
that contribute to cross-plane phonon transport at each frequency
$\omega$. This problem is particularly tricky for a single-layer
2D crystal since it has no extended volume in the cross-plane direction
as shown in Fig.~\ref{fig:DiffuseTransmissonSchematics}(a). Instead,
we \emph{estimate} $M_{\text{2D}}(\omega)$ in our model phenomenologically
by assuming that the interface between the 2D crystal and a substrate
consisting of a semi-infinite number of the same 2D material is acoustically
transparent, e.g. a single-layer MoS$_{2}$ (1L-MoS$_{2}$) on a bulk
MoS$_{2}$ ($\lim_{N\rightarrow\infty}N$L-MoS$_{2}$) substrate,
such that the transmission function between the two materials is equal
to the number of transmitting modes in the single-layer 2D crystal.
To do this, we make the replacement $K\rightarrow K_{\text{2D}}$
and $D_{\text{sub}}(\boldsymbol{q},\omega)\rightarrow D_{\text{2D},\infty}(\boldsymbol{q},\omega)$
in Eq.~(\ref{eq:Transmission_coherent}), where 
\begin{equation}
D_{\text{\text{2D},\ensuremath{\infty}}}(\boldsymbol{q},\omega)=\Theta(z)D_{+}(\boldsymbol{q},\omega)+\Theta(-z)D_{-}(\boldsymbol{q},\omega)\label{eq:SurfaceResponseLayeredSubstrate}
\end{equation}
is the surface response function of a semi-infinite 2D layered substrate~\citep{ZYOng:Manuscript20_BoronNitride},
with
\[
D_{\pm}(\boldsymbol{q},\omega)=\frac{2}{z(\boldsymbol{q},\omega)\pm\sqrt{z(\boldsymbol{q},\omega)^{2}-4z(\boldsymbol{q},\omega)K_{\text{2D}}}}
\]
for $z(\boldsymbol{q},\omega)=\lim_{\gamma\rightarrow0}D_{\text{2D}}(\boldsymbol{q},\omega)^{-1}=\rho\omega^{2}-\kappa q^{4}$,
and $K_{\text{2D}}$ is the interlayer spring constant in the substrate.
Hence, we obtain the expression analogous to Eq.~(\ref{eq:Transmission_coherent}),
i.e., 
\begin{equation}
M_{\text{2D}}(\omega)=\frac{1}{(2\pi)^{2}}\int_{q<q_{\text{max}}}d^{2}q\frac{4K_{\text{2D}}^{2}\text{Im}D_{\text{2D},\infty}(\boldsymbol{q},\omega)\text{Im}D_{\text{2D}}(\boldsymbol{q},\omega)}{|1-K_{\text{2D}}[D_{\text{2D},\infty}(\boldsymbol{q},\omega)+D_{\text{2D}}(\boldsymbol{q},\omega)]|^{2}}\label{eq:TMD_Ballistic_Transmission}
\end{equation}
which can be evaluated numerically. The set up of the calculation
for $M_{\text{2D}}(\omega)$ for a single-layer TMD (1L-TMD) is shown
in Fig.~\ref{fig:DiffuseTransmissonSchematics}(b). In the case of
SiO$_{2}$ as shown in Fig.~\ref{fig:DiffuseTransmissonSchematics}(c),
we estimate the number of modes per unit area in an isotropic elastic
substrate $M_{\text{sub}}(\omega)$ as
\begin{equation}
M_{\text{sub}}(\omega)=\frac{\omega^{2}}{4\pi}\left(\frac{1}{c_{L}^{2}}+\frac{2}{c_{T}^{2}}\right)\Theta(\omega_{D}-\omega)\ .\label{eq:Substrate_Ballistic_Transmission}
\end{equation}
Since $M_{\text{2D}}(\omega)\ll M_{\text{sub}}(\omega)$ for $\omega<\omega_{D}$,
we have $\Xi_{\text{diff}}(\omega)\approx M_{\text{2D}}(\omega)\Theta(\omega_{D}-\omega)$
and Eq.~(\ref{eq:TBC_diffuse_limit}) becomes
\begin{equation}
G_{\text{ph}}^{\text{diff}}\approx\frac{1}{2\pi}\int_{0}^{\omega_{D}}d\omega\ \hbar\omega\frac{dN(\omega,T)}{dT}M_{\text{2D}}(\omega)\label{eq:TBC_diffuse_limit_approx}
\end{equation}
which depends on the substrate through its longitudinal Debye frequency
$\omega_{D}$ and the 2D crystal though $M_{\text{2D}}(\omega)$.
The expression in Eq.~(\ref{eq:TBC_diffuse_limit_approx}) suggests
that in the diffuse limit, the phononic TBC for a particular 2D crystal
can be maximized only through $\omega_{D}$ which depends on $c_{L}$.
Hence, the phononic TBC would be high for hard insulators such as
diamond and sapphire for which $c_{L}=17500$ and $10890$ m/s, respectively~\citep{ETSwartz:RMP89_Thermal}.

\begin{table}
\begin{centering}
\begin{tabular}{|c|c|c|}
\hline 
TMD & WS$_{2}$ & MoS$_{2}$\tabularnewline
\hline 
\hline 
$K$ ($10^{19}$ Nm$^{-3}$) & $6.12$ & $4.94$\tabularnewline
\hline 
$\kappa$ (eV) & $11.25$~\citep{KLai:JPDAP16_Bending} & $9.61$~\citep{KLai:JPDAP16_Bending}\tabularnewline
\hline 
$\rho$ ($10^{-7}$ kgm$^{-2}$) & $47.9$ & $31.3$\tabularnewline
\hline 
$\alpha$ & \multicolumn{2}{c|}{$100$~\citep{ZYOng:PRB16_Theory}}\tabularnewline
\hline 
$\rho_{\text{sub}}$ (kgm$^{-3}$) & \multicolumn{2}{c|}{$2200$~\citep{BNJPersson:JPCM11_Phononic}}\tabularnewline
\hline 
$c_{L}$ (ms$^{-1}$) & \multicolumn{2}{c|}{$5953$~\citep{BNJPersson:JPCM11_Phononic}}\tabularnewline
\hline 
$c_{T}$ (ms$^{-1}$) & \multicolumn{2}{c|}{$3743$~\citep{BNJPersson:JPCM11_Phononic}}\tabularnewline
\hline 
$\omega_{D}$ (meV) & \multicolumn{2}{c|}{$61.8$}\tabularnewline
\hline 
$K_{\text{2D}}$ ($10^{19}$ Nm$^{-3}$) & $9.55$~\citep{LLiang:ACSNano17_LowFrequency} & $8.90$~\citep{LLiang:ACSNano17_LowFrequency}\tabularnewline
\hline 
\end{tabular}
\par\end{centering}
\caption{Parameters in our numerical simulations of Eqs.~(\ref{eq:CoherentPhononTBC})
and (\ref{eq:TBC_diffuse_limit}). $K$, which depends on the type
of 2D crystal, is the spring constant per unit area for the OH-terminated
SiO$_{2}$ interface and is calculated using DFT~\citep{ZYOng:PRB16_Theory}.
$\kappa$ and $\rho$ are respectively the intrinsic bending rigidity
and mass density per unit area of the 2D crystal used in Eq.~(\ref{Eq:2DFlexuralGreensFunction}).
$K_{\text{2D}}$ is the spring constant per unit area for the bulk
version of the 2D crystal~\citep{LLiang:ACSNano17_LowFrequency}.}
\label{Tab:PhononSimParameters}
\end{table}

\subsection{Heat dissipation by remote phonon scattering}

To model the electronic TBC $G_{\text{el}}$, we use the theory developed
in Ref.~\citep{ZYOng:Manuscript20_Remote}. The expression for $G_{\text{el}}$
is written as a 2D integral that can be evaluated numerically, i.e.,
\begin{equation}
G_{\text{el}}=\sum_{\gamma=\text{SO1},\text{SO2}}\mathcal{F}_{\gamma}\int_{0}^{q_{\text{max}}}dq\ \frac{\exp(-2qd)}{\varepsilon(q)}\int_{0}^{\infty}d\mu^{\prime}\frac{\text{Im}\mathcal{P}(\boldsymbol{q},\omega_{\gamma};\mu^{\prime},0)}{4k_{B}T\cosh^{2}(\frac{\mu-\mu^{\prime}}{2k_{B}T})}\label{eq:TBCExpression}
\end{equation}
where $q_{\text{max}}=2\times10^{10}$ m$^{-1}$ is the cutoff wave
vector like in Eq.~(\ref{eq:Transmission_coherent}), $\omega_{\gamma}$
is the frequency of the $\gamma$ phonon, $d$ is the TMD-substrate
gap size, $\mu$ is the chemical potential, and 
\begin{equation}
\mathcal{F}_{\gamma}=\frac{e^{2}\hbar^{2}\omega_{\gamma}^{3}}{\pi k_{B}T^{2}}[N(\omega_{\gamma},T)+1]N(\omega_{\gamma},T)\left(\frac{1}{\epsilon_{\gamma,\text{hi}}}-\frac{1}{\epsilon_{\gamma,\text{lo}}}\right)\ .\label{eq:FCoefficient}
\end{equation}
In Eq.~(\ref{eq:TBCExpression}), $\mathcal{P}(\boldsymbol{q},\omega_{\gamma};\mu,T)$
is the electron polarizability~\citep{ZYOng:Manuscript20_Remote}
while the screening function $\varepsilon(q)$ is given by~\citep{ZYOng:PRB12_Charged}
\begin{equation}
\varepsilon(q)^{-1}=1+\frac{e^{2}\text{Re}\mathcal{P}(\boldsymbol{q},0;\mu,T)}{2\epsilon_{0}q}\left[1-\frac{\epsilon_{\text{sub}}^{\infty}-\epsilon_{0}}{\epsilon_{\text{sub}}^{\infty}+\epsilon_{0}}\exp(-2qd)\right]\label{eq:ScreeningFunction}
\end{equation}
where $e$ and $\epsilon_{0}$ are the electron charge and the permittivity
of vacuum, respectively. In Eq.~(\ref{eq:FCoefficient}), the expressions
for $\epsilon_{\text{SO1}\gamma,\text{hi}}$, $\epsilon_{\text{SO1},\text{lo}}$,
$\epsilon_{\text{SO2},\text{hi}}$ and $\epsilon_{\text{SO2},\text{lo}}$
are given by $\epsilon_{\text{SO1},\text{hi}}=\frac{1}{2}\left[\epsilon_{\text{sub}}^{\infty}\left(\frac{\omega_{\text{LO2}}^{2}-\omega_{\text{SO1}}^{2}}{\omega_{\text{TO2}}^{2}-\omega_{\text{SO1}}^{2}}\right)+\epsilon_{0}\right]$,
$\epsilon_{\text{SO1},\text{lo}}=\frac{1}{2}\left[\epsilon_{\text{sub}}^{\infty}\left(\frac{\omega_{\text{LO1}}^{2}}{\omega_{\text{TO1}}^{2}}\right)\left(\frac{\omega_{\text{LO2}}^{2}-\omega_{\text{SO1}}^{2}}{\omega_{\text{TO2}}^{2}-\omega_{\text{SO1}}^{2}}\right)+\epsilon_{0}\right]$,
$\epsilon_{\text{SO2},\text{hi}}=\frac{1}{2}\left[\epsilon_{\text{sub}}^{\infty}\left(\frac{\omega_{\text{LO1}}^{2}-\omega_{\text{SO2}}^{2}}{\omega_{\text{TO1}}^{2}-\omega_{\text{SO2}}^{2}}\right)+\epsilon_{0}\right]$
and $\epsilon_{\text{SO2},\text{lo}}=\frac{1}{2}\left[\epsilon_{\text{sub}}^{\infty}\left(\frac{\omega_{\text{LO2}}^{2}}{\omega_{\text{TO2}}^{2}}\right)\left(\frac{\omega_{\text{LO1}}^{2}-\omega_{\text{SO2}}^{2}}{\omega_{\text{TO1}}^{2}-\omega_{\text{SO2}}^{2}}\right)+\epsilon_{0}\right]$
where $\omega_{\text{TO1}}$ and $\omega_{\text{TO2}}$ are the transverse
optical phonon frequencies associated with the bulk polar optical
phonons of the substrate.

We use the simulation parameters from Table~\ref{Tab:RemotePhononParameters}
for our $G_{\text{el}}$ calculations.  The longitudinal optical
(LO) phonon frequencies $\omega_{\text{LO1}}$ and $\omega_{\text{LO2}}$
are determined from the zeros of $\epsilon_{\text{sub}}(\omega)=\epsilon_{\text{sub}}^{\infty}+(\epsilon_{\text{sub}}^{i}-\epsilon_{\text{sub}}^{\infty})\frac{\omega_{\text{TO2}}^{2}}{\omega^{2}-\omega_{\text{TO2}}^{2}}+(\epsilon_{\text{sub}}^{0}-\epsilon_{\text{sub}}^{i})\frac{\omega_{\text{TO1}}^{2}}{\omega^{2}-\omega_{\text{TO1}}^{2}}$
while the surface optical (SO) phonon frequencies $\omega_{\text{SO1}}$
and $\omega_{\text{SO2}}$ are determined from solving $\epsilon_{\text{sub}}(\omega)+\epsilon_{0}=0$~\citep{AKonar:PRB10_Effect}.

\begin{table}
\begin{centering}
\begin{tabular}{|c|c|c|}
\hline 
Substrate & WS$_{2}$ & MoS$_{2}$\tabularnewline
\hline 
\hline 
$m_{e}/m_{0}$ & $0.31$ & $0.51$ \tabularnewline
\hline 
$d$ ($\text{Å}$) & \multicolumn{2}{c|}{3.0~\citep{ZYOng:Manuscript20_Remote}}\tabularnewline
\hline 
$g_{s}$  & \multicolumn{2}{c|}{2}\tabularnewline
\hline 
$g_{v}$  & \multicolumn{2}{c|}{2}\tabularnewline
\hline 
$\epsilon_{\text{sub}}^{\infty}/\epsilon_{0}$ & \multicolumn{2}{c|}{2.50}\tabularnewline
\hline 
$\epsilon_{\text{sub}}^{i}/\epsilon_{0}$ & \multicolumn{2}{c|}{3.05}\tabularnewline
\hline 
$\epsilon_{\text{sub}}^{0}/\epsilon_{0}$ & \multicolumn{2}{c|}{3.90}\tabularnewline
\hline 
$\omega_{\text{TO1}}$ (meV) & \multicolumn{2}{c|}{55.60}\tabularnewline
\hline 
$\omega_{\text{TO2}}$ (meV) & \multicolumn{2}{c|}{138.10}\tabularnewline
\hline 
$\omega_{\text{SO1}}$ (meV) & \multicolumn{2}{c|}{60.99}\tabularnewline
\hline 
$\omega_{\text{SO2}}$ (meV) & \multicolumn{2}{c|}{148.97}\tabularnewline
\hline 
\end{tabular}
\par\end{centering}

\caption{Remote phonon scattering simulation parameters for WS$_{2}$ and MoS$_{2}$~\citep{ZJin:PRB14_Intrinsic}.
The effective electron masses $m_{e}$ are expressed in terms of the
free electron mass $m_{0}$ and taken from Ref.~\citep{ZJin:PRB14_Intrinsic}.
The variables $m_{e}$, $g_{s}$ and $g_{v}$ are used in $\mathcal{P}(\boldsymbol{q},\omega_{\gamma};\mu,T)$.
The parameters $\epsilon_{\text{sub}}^{\infty}$, $\epsilon_{\text{sub}}^{i}$,
$\epsilon_{\text{sub}}^{0}$, $\omega_{\text{TO1}}$ and $\omega_{\text{TO2}}$
for SiO$_{2}$ are taken from Ref.~\citep{ZYOng:PRB12_Theory}.}

\label{Tab:RemotePhononParameters}
\end{table}

\section{Numerical results and discussion}

\subsection{Comparison of coherent and diffuse phononic TBC}

\begin{figure}
\begin{centering}
\includegraphics[scale=0.5]{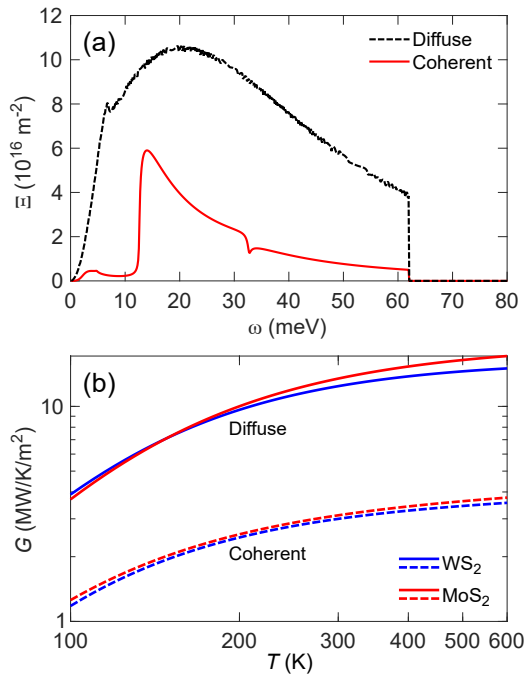}
\par\end{centering}
\caption{(a) Coherent and diffuse phonon transmission spectra for the MoS$_{2}$-SiO$_{2}$
interface at $300$ K from Eqs.~(\ref{eq:Transmission_coherent})
and (\ref{eq:Transmission_DMM}), respectively. (b) Temperature dependence
of the coherent (dotted lines) and the diffuse (solid lines) phononic
TBC ($G_{\text{ph}}^{\text{coh}}$ vs. $G_{\text{ph}}^{\text{diff}}$)
for the WS$_{2}$-SiO$_{2}$ (blue lines) and MoS$_{2}$-SiO$_{2}$
(red lines) interface from $T=100$ to $600$ K. At 300 K, we have
$G_{\text{ph}}^{\text{coh}}=3.1$ MW/K/m$^{2}$ and $G_{\text{ph}}^{\text{diff}}=13.5$
MW/K/m$^{2}$ ($G_{\text{ph}}^{\text{coh}}=3.0$ MW/K/m$^{2}$ and
$G_{\text{ph}}^{\text{diff}}=12.4$ MW/K/m$^{2}$) for the MoS$_{2}$-SiO$_{2}$
(WS$_{2}$-SiO$_{2}$) interface.}
\label{fig:PhononTBC}
\end{figure}

It is intuitively expected that disorder in the interface leads to
higher thermal resistance because of increased phonon scattering~\citep{PHopkins:ISRNME13_Thermal}.
However, the presence of interfacial disorder can enlarge the scattering
phase space and possibly improve interfacial thermal transport by
allowing energy transfer to proceed through scattering pathways that
do not conserve parallel momentum~\citep{ZTian:PRB12_Enhancing}.
This difference in scattering is manifested in the transmission spectra
$\Xi_{\text{coh}}(\omega)$ and $\Xi_{\text{diff}}(\omega)$ from
Eqs.~(\ref{eq:Transmission_coherent}) and (\ref{eq:Transmission_DMM})
in Fig.~\ref{fig:PhononTBC}(a) which shows a marked difference in
the transmission spectra for the MoS$_{2}$-SiO$_{2}$ interface,
especially at low frequencies ($\omega<12$ meV) where $\Xi_{\text{diff}}(\omega)$
is substantially higher than $\Xi_{\text{coh}}(\omega)$. The sharp
drop in the transmission spectra observed at $61.8$ meV is due to
the frequency cutoff at the longitudinal Debye frequency of the substrate. 

To understand the implications of the difference in the transmission
spectra between the ordered and disordered TMD-substrate interface,
we compare $G_{\text{ph}}^{\text{coh}}$ and $G_{\text{ph}}^{\text{diff}}$
of Eqs .~(\ref{eq:CoherentPhononTBC}) and (\ref{eq:TBC_diffuse_limit})
from $T=100$ to $600$ K, using the parameters in Table~\ref{Tab:PhononSimParameters},
for the WS$_{2}$-SiO$_{2}$ and MoS$_{2}$-SiO$_{2}$ interface in
Fig.~\ref{fig:PhononTBC}(b). For the MoS$_{2}$-SiO$_{2}$ interface
at $T=300$ K, we obtain $G_{\text{ph}}^{\text{diff}}=13.5$ MW/K/m$^{2}$,
which is significantly larger than $G_{\text{ph}}^{\text{coh}}=3.1$
MW/K/m$^{2}$ and more comparable to the value of $G=14.0\pm3.7$
MW/K/m$^{2}$ at $311$ K in Ref.~\citep{EYalon:ACSApplMaterInterf17_Temperature},
$G=20.3$\textendash $33.5$ MW/K/m$^{2}$ at $295$ K in Ref.~\citep{PYasaei:AdvMaterInterf17_Interfacial},
$G=21.0$ MW/K/m$^{2}$ at $300$ K in Ref.~\citep{JGuo:JPDAP19_Conformal}
and $G=18.6$ MW/K/m$^{2}$ at $300$ K in Ref.~\citep{YYu:PRAppl20_InPlane}.
This suggests that the MoS$_{2}$-SiO$_{2}$ interface is disordered,
possibly connected to the absence of long-range order in a-SiO$_{2}$
and chemical inhomogeneity of its surface~\citep{DRhodes:NatMater19_Disorder},
and that its thermal transport properties are better described by
the diffuse phonon model. The larger $G_{\text{ph}}^{\text{diff}}$
for the MoS$_{2}$-SiO$_{2}$ interface is also closer to classical
MD simulation results ($G=15.6$ MW/K/m$^{2}$ in Ref.~\citep{PYasaei:AdvMaterInterf17_Interfacial},
$G=12.2$\textendash $23.5$ MW/K/m$^{2}$ in Ref.~\citep{HFarahani:CMS18_Interfacial}
and $G=25.6\pm3.3$ MW/K/m$^{2}$ in Ref.~\citep{SVSuryavanshi:JAP19_Thermal}).
For the WS$_{2}$-SiO$_{2}$ interface at 300 K, we also find that
$G_{\text{ph}}^{\text{diff}}=12.4$ MW/K/m$^{2}$ is significantly
larger than $G_{\text{ph}}^{\text{coh}}=3.0$ MW/K/m$^{2}$. The good
agreement of $G_{\text{ph}}^{\text{diff}}$ with experimentally derived
values for the MoS$_{2}$-SiO$_{2}$ interface suggests that the real
TBC value for the WS$_{2}$-SiO$_{2}$ interface is probably closer
to the predicted $G_{\text{ph}}^{\text{diff}}=12.4$ MW/K/m$^{2}$.
We also note that the $G_{\text{ph}}^{\text{coh}}$ and $G_{\text{ph}}^{\text{diff}}$
values for the WS$_{2}$-SiO$_{2}$ and MoS$_{2}$-SiO$_{2}$ interface
are comparable because of their similar elasticity parameters ($\rho$,
$\kappa$ and $K_{\text{2D}}$). 

\subsection{Interpretation of experimental data for MoS$_{2}$-SiO$_{2}$ interface}

\begin{figure}
\begin{centering}
\includegraphics[scale=0.5]{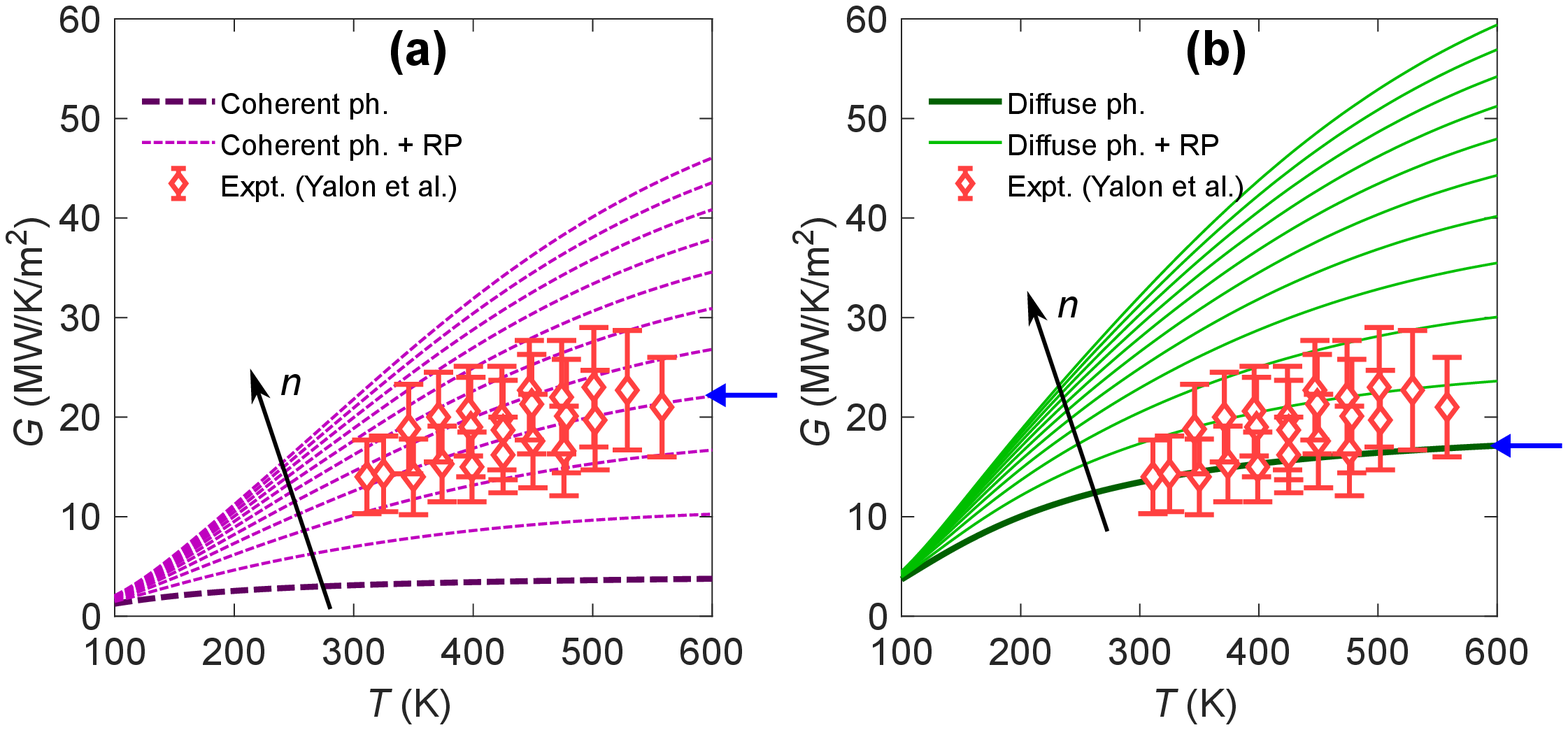}
\par\end{centering}
\caption{Temperature dependence of (a) $G_{\text{ph}}^{\text{coh}}$ (thick
dashed line labeled `Coherent ph.')  and $G_{\text{ph}}^{\text{coh}}+G_{\text{el}}$
(fine purple dashed lines labeled `Coherent ph. + RP'), and (b) $G_{\text{ph}}^{\text{diff}}$
(thick solid line labeled `Diffuse ph.') and $G_{\text{ph}}^{\text{diff}}+G_{\text{el}}$
(fine green solid lines labeled `Diffuse ph. + RP') for the MoS$_{2}$-SiO$_{2}$
interface at different values of the electron density between $n=10^{11}$
to $10^{12}$ cm$^{-2}$ in steps of $\Delta n=10^{11}$ cm$^{-2}$.
As $n$ increases, so do $G_{\text{ph}}^{\text{coh}}+G_{\text{el}}$
and $G_{\text{ph}}^{\text{diff}}+G_{\text{el}}$ as indicated by the
black arrow. The corresponding experimental TBC data from Yalon et
al.~\citep{EYalon:ACSApplMaterInterf17_Temperature} are indicated
by red diamond symbols with error bars. The $G_{\text{ph}}^{\text{coh}}+G_{\text{el}}$
($n=3\times10^{11}$ cm$^{-2}$) and $G_{\text{ph}}^{\text{diff}}+G_{\text{el}}$
($n=0$ cm$^{-2}$) curves that best fit the experimental data are
indicated by the short blue arrows.}

\label{fig:YalonTBCDataAnalysis}
\end{figure}

We combine the results for $G_{\text{el}}$, $G_{\text{ph}}^{\text{coh}}$
and $G_{\text{ph}}^{\text{diff}}$ to analyze the experimental TBC
data ($G_{\text{expt}}$) for the MoS$_{2}$-SiO$_{2}$ interface
from Yalon et al.~\citep{EYalon:ACSApplMaterInterf17_Temperature}.
As mentioned earlier, we assume that the phononic and electronic TBC
components are independent and can be added in parallel, and that
the electronic TBC is unaffected by the disorder at the TMD-substrate
interface. In Fig.~\ref{fig:YalonTBCDataAnalysis}, we plot $G_{\text{ph}}^{\text{coh}}$,
$G_{\text{ph}}^{\text{diff}}$, $G_{\text{ph}}^{\text{coh}}+G_{\text{el}}$
(coherent phononic and electronic) and $G_{\text{ph}}^{\text{diff}}+G_{\text{el}}$
(diffuse phononic and electronic) at different values of the electron
density from $n=10^{11}$ to $10^{12}$ cm$^{-2}$ in steps of $\Delta n=10^{11}$
cm$^{-2}$ as a function of temperature and compare them to the $G_{\text{expt}}$
data. 

At $n=0$ cm$^{-2}$, there is no electronic contribution to the overall
TBC and the theoretical TBC values are given by $G_{\text{ph}}^{\text{coh}}$
or $G_{\text{ph}}^{\text{diff}}$ which we can treat as the baseline
TBC. We find that the $G_{\text{expt}}$ values are significantly
higher than $G_{\text{ph}}^{\text{coh}}$ and in much closer agreement
with $G_{\text{ph}}^{\text{diff}}$ over the temperature range of
311 to 558 K. This suggests that the diffuse phonon model captures
the essential physics of heat dissipation at the MoS$_{2}$-SiO$_{2}$
interface and that electronic contribution to the TBC is not reflected
in the $G_{\text{expt}}$ data.

Nonetheless, we explore the effects of the electronic contribution
to the total TBC which can be significant. Figure~\ref{fig:YalonTBCDataAnalysis}
shows that $G_{\text{expt}}$ data can also be fitted by $G_{\text{ph}}^{\text{coh}}+G_{\text{el}}$
at $n=0.3\times10^{12}$ cm$^{-2}$, i.e., the discrepancies between
the experimental TBC data and the predicted coherent phononic TBC
ones can be eliminated by adding the contribution from the remote
phonon scattering of TMD electrons. Although the predictions for $G_{\text{ph}}^{\text{coh}}+G_{\text{el}}$
at the relatively low electron density of $n=0.3\times10^{12}$ cm$^{-2}$
seem to fit the $G_{\text{expt}}$ data slightly better at higher
temperatures that the $G_{\text{ph}}^{\text{diff}}$ predictions do,
the error bars in the $G_{\text{expt}}$ data are too large for us
to exclude $G_{\text{ph}}^{\text{coh}}$ or $G_{\text{ph}}^{\text{diff}}$
as the phononic component of the total TBC. Nonetheless, we favor
$G_{\text{ph}}^{\text{diff}}$ as the phononic component because its
predicted values are closer to classical MD simulation results~\citep{PYasaei:AdvMaterInterf17_Interfacial,HFarahani:CMS18_Interfacial,SVSuryavanshi:JAP19_Thermal}
which have no electronic contribution. 

The substantial spread in the $G_{\text{ph}}^{\text{diff}}+G_{\text{el}}$
predictions in the $n=0$ to $10^{12}$ cm$^{-2}$ range may also
explain the variation in the experimental TBC values for the MoS$_{2}$-SiO$_{2}$
interface. It has been reported that the intrinsic doping of single-layer
MoS$_{2}$ can reach up to $\sim10^{13}$ cm$^{-2}$ on SiO$_{2}$
substrates~\citep{BBaugher:NL13_Intrinsic,KMak:NatMater13_Tightly}
although not all the electrons are delocalized~\citep{HQiu:NatCommun13_Hopping}.
In samples grown by chemical vapor deposition, a maximum electron
density of $n\approx10^{13}$ cm$^{-2}$ can be reached using an applied
gate voltage~\citep{KSmithe:2DM17_Intrinsic}. Nonetheless, we limit
our analysis in the following discussion to a conservative $n\leq10^{12}$
cm$^{-2}$ range as there has not been any systematic attempt to measure
how the TBC varies with $n$. As $n$ increases from $0$ to $10^{12}$
cm$^{-2}$, the theoretical $G_{\text{ph}}^{\text{diff}}+G_{\text{el}}$
TBC at $300$ K increases from $13.5$ to $32.1$ MW/K/m$^{2}$, producing
a spread comparable to the TBC variation ($G=20.3$ to $33.5$ MW/K/m$^{2}$)
obtained at $295$ K in Ref.~\citep{PYasaei:AdvMaterInterf17_Interfacial}.
This suggests that the spread in TBC values may be caused by the variability
in the electron density due to intrinsic doping in MoS$_{2}$~\citep{WLeong:Nanoscale15_Tuning}
which is often $n$-doped as a result of sulfur atom vacancies~\citep{HQiu:NatCommun13_Hopping}.
It also implies that it should be possible to increase the TBC of
the MoS$_{2}$-SiO$_{2}$ interface by doping the MoS$_{2}$ through
the introduction of dopants~\citep{DKiriya:JACS14_Air,PRastogi:JPCC14_Doping}
or the use of a metal gate electrode (as in a field-effect transistor)
to modulate the electron density. In addition to the change in the
TBC as $n$ increases, another striking feature is the change in the
temperature dependence of the TBC. 

\subsection{Phononic and electronic TBC for WS$_{2}$-SiO$_{2}$ interface}

\begin{figure}
\begin{centering}
\includegraphics[scale=0.5]{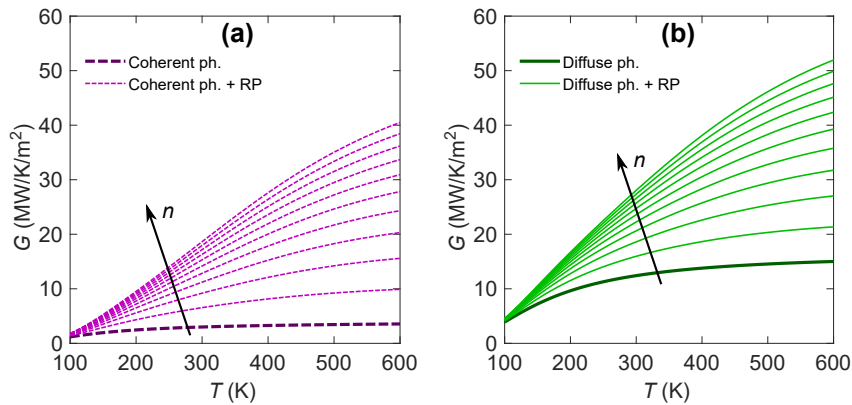}
\par\end{centering}
\caption{Temperature dependence of (a) $G_{\text{ph}}^{\text{coh}}$ (thick
dashed line labeled `Coherent ph.')  and $G_{\text{ph}}^{\text{coh}}+G_{\text{el}}$
(fine purple dashed lines labeled `Coherent ph. + RP') and (b) $G_{\text{ph}}^{\text{diff}}$
(thick solid line labeled `Diffuse ph.') and $G_{\text{ph}}^{\text{diff}}+G_{\text{el}}$
(fine green solid lines labeled `Diffuse ph. + RP') for the WS$_{2}$-SiO$_{2}$
interface at different values of the electron density between $n=10^{11}$
to $10^{12}$ cm$^{-2}$ in steps of $\Delta n=10^{11}$ cm$^{-2}$.
As $n$ increases, so do $G_{\text{ph}}^{\text{coh}}+G_{\text{el}}$
and $G_{\text{ph}}^{\text{diff}}+G_{\text{el}}$ as indicated by the
black arrow. }

\label{fig:WS2_results}
\end{figure}

We also present the simulated electronic and phononic TBC for the
WS$_{2}$-SiO$_{2}$ interface in Fig.~\ref{fig:WS2_results} in
which we plot $G_{\text{ph}}^{\text{coh}}$, $G_{\text{ph}}^{\text{diff}}$,
$G_{\text{ph}}^{\text{coh}}+G_{\text{el}}$ (coherent phononic and
electronic) and $G_{\text{ph}}^{\text{diff}}+G_{\text{el}}$ (diffuse
phononic and electronic) at different values of the electron density
from $n=10^{11}$ to $10^{12}$ cm$^{-2}$ in steps of $\Delta n=10^{11}$
cm$^{-2}$ as a function of temperature. Unlike Fig.~\ref{fig:YalonTBCDataAnalysis},
no comparison with experimental data is made because no such data
is available for the WS$_{2}$-SiO$_{2}$ interface. Nevertheless,
the data in Fig.~\ref{fig:WS2_results} can be useful for comparison
with future experimental TBC measurements of the WS$_{2}$-SiO$_{2}$
interface and understanding its TBC variability.

We observe similar trends in Fig.~\ref{fig:WS2_results} to the data
in Fig.~\ref{fig:YalonTBCDataAnalysis}. At 300 K, as $n$ increases
from $0$ to $10^{12}$ cm$^{-2}$, $G_{\text{ph}}^{\text{coh}}+G_{\text{el}}$
increases from $3.0$ to $18.6$ MW/K/m$^{2}$ while $G_{\text{ph}}^{\text{diff}}+G_{\text{el}}$
increases from $12.4$ to $28.1$ MW/K/m$^{2}$. We note that the
$G_{\text{ph}}^{\text{coh}}$, $G_{\text{ph}}^{\text{diff}}$ and
$G_{\text{el}}$ data for the WS$_{2}$-SiO$_{2}$ interface are comparable
but smaller than their corresponding values for the MoS$_{2}$-SiO$_{2}$
interface in Fig.~\ref{fig:YalonTBCDataAnalysis}. The large change
in the $G_{\text{ph}}^{\text{coh}}+G_{\text{el}}$ and $G_{\text{ph}}^{\text{diff}}+G_{\text{el}}$
data with respect to $n$ means that a change in electron density,
from either intrinsic doping or an applied gate voltage, can also
lead to a significant detectable change in the measured TBC as with
the MoS$_{2}$-SiO$_{2}$ interface.

\section{Summary and conclusion}

In this work, we analyze the theoretical phononic and electronic TBC
of the WS$_{2}$-SiO$_{2}$ and MoS$_{2}$-SiO$_{2}$ interface. To
describe the phononic TBC contribution for the disordered TMD-substrate
interface, we introduce a diffuse phonon model. We compare the coherent
and diffuse phonon models for the WS$_{2}$-SiO$_{2}$ and MoS$_{2}$-SiO$_{2}$
interface, and find that the diffuse phonon model yields significantly
higher TBC values that fit the TBC data for the MoS$_{2}$-SiO$_{2}$
interface from molecular dynamics simulations and thermometric experiments
better, implying that the TMD-substrate interfaces are disordered.
Our analysis of the experimental TBC data from Ref.~\citep{EYalon:ACSApplMaterInterf17_Temperature}
for the MoS$_{2}$-SiO$_{2}$ interface indicates that the overall
TBC is dominated by diffuse phonon transport although the electronic
contribution from remote phonon scattering can be significant even
at low electron densities and becomes comparable to the phononic contribution
at higher electron densities. The simulated phononic and electronic
TBC data for the WS$_{2}$-SiO$_{2}$ interface also indicate that
the electronic contribution to its TBC is also significant. Our results
show that the spread in experimental TBC values of the MoS$_{2}$-SiO$_{2}$
interface can possibly be explained by the variability in intrinsic
doping which affects the electronic TBC. 
\begin{acknowledgments}
We gratefully acknowledge support from the Science and Engineering
Research Council through grant (152-70-00017) and use of computing
resources at the A{*}STAR Computational Resource Centre and National
Supercomputer Centre, Singapore. We also thank Eilam Yalon (Technion,
Israel Institute of Technology) for sharing the experimental data
from Ref.~\citep{EYalon:ACSApplMaterInterf17_Temperature} with us.
\end{acknowledgments}

\bibliographystyle{apsrev4-1}
\bibliography{PaperReferences,LocalReferences}

\end{document}